\documentclass{IEEEtran}

\usepackage{cite}
\usepackage{amsmath,amssymb,amsfonts}
\usepackage{algorithmic}
\usepackage{graphicx}
\usepackage{textcomp}

\usepackage{lipsum}
\usepackage{multicol}
\usepackage{multirow}

\usepackage{xcolor}

\usepackage{array}
\newcommand{\PreserveBackslash}[1]{\let\temp=\\#1\let\\=\temp}
\newcolumntype{C}[1]{>{\PreserveBackslash\centering}p{#1}}
\newcolumntype{R}[1]{>{\PreserveBackslash\raggedleft}p{#1}}
\newcolumntype{L}[1]{>{\PreserveBackslash\raggedright}p{#1}}

\newcommand{\Fp}{\mathbb{F}_p}
\newcommand{\Fq}{\mathbb{F}_q}
\newcommand{\Fpa}{\mathbb{F}_{p^{2}}}

\newcommand{\Fpc}{\mathbb{F}_{p^{6}}}
\newcommand{\Fpd}{\mathbb{F}_{p^{12}}}

\newcommand{\Ga}{\mathbb{G}_1}
\newcommand{\Gb}{\mathbb{G}_2}
\newcommand{\Gt}{\mathbb{G}_{T}}

\usepackage{hyperref}
\hypersetup{
colorlinks=true,
linkcolor=blue,
filecolor=magenta,      
urlcolor=blue,
}

\def\BibTeX{{\rm B\kern-.05em{\sc i\kern-.025em b}\kern-.08em
    T\kern-.1667em\lower.7ex\hbox{E}\kern-.125emX}}
\begin{document}
\title{ A Low-Power BLS12-381 Pairing Crypto-Processor for Internet-of-Things Security Applications }

\author{Utsav Banerjee, \IEEEmembership{Member, IEEE}, and Anantha P. Chandrakasan, \IEEEmembership{Fellow, IEEE}
%\thanks{Manuscript received June 30, 2021; revised September 27, 2021; accepted October 25, 2021. This work was supported by Texas Instruments.}
%\thanks{}
%\thanks{}
%\thanks{}
\thanks{The authors are with the Department of Electrical Engineering and Computer Science, Massachusetts Institute of Technology, Cambridge, MA 02139 USA (e-mail: utsav@alum.mit.edu).}
%\thanks{Digital Object Identifier}
\thanks{\textcopyright $\,$ 2021 IEEE. Personal use of this material is permitted. Permission from IEEE must be obtained for all other uses, in any current or future media, including reprinting/republishing this material for advertising or promotional purposes, creating new collective works, for resale or redistribution to servers or lists, or reuse of any copyrighted component of this work in other works.}
\thanks{A revised version of this paper was published in the IEEE Solid-State Circuits Letters (SSC-L) - DOI: \href{https://dx.doi.org/10.1109/LSSC.2021.3124074}{10.1109/LSSC.2021.3124074}}
}

\maketitle

\begin{abstract}
We present the first BLS12-381 elliptic curve pairing crypto-processor for Internet-of-Things (IoT) security applications. Efficient finite field arithmetic and algorithm-architecture co-optimizations together enable two orders of magnitude energy savings. We implement several countermeasures against timing and power side-channel attacks. Our crypto-processor is programmable to provide the flexibility to accelerate various elliptic curve and pairing-based protocols such as signature aggregation and functional encryption.
\end{abstract}


\section{Introduction}

\begin{figure}[!t]
\centering
\includegraphics[width=3.4in]{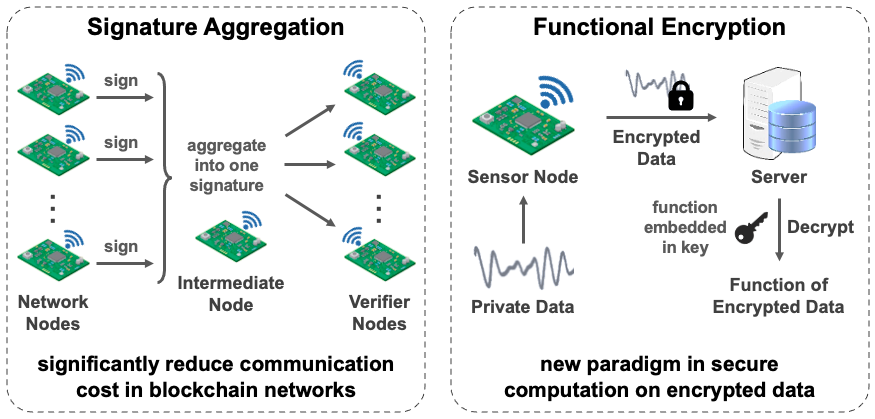}
\caption{Two important applications of pairing-based cryptography.}
\label{pbc_applications}
\end{figure}

Elliptic curves are used as the de facto standard for traditional public key cryptography such as key establishment, digital signatures, authenticated key exchange and public key encryption \cite{hankerson_ecc_2006}.
Pairing-based cryptography (PBC), a variant of elliptic curve cryptography (ECC), uses bilinear maps between elliptic curves and finite fields to enable novel applications beyond traditional key exchange and signatures \cite{mrabet_pairing_2017}.
Fig. \ref{pbc_applications} shows two such applications – (a) signature aggregation, where arbitrarily large number of signatures are compressed into one to resolve communication bottleneck in mesh networks such as blockchain \cite{boneh_blockchain_2018}, and (b) functional encryption, which allows computing on encrypted data with a function embedded in the decryption key. In particular, pairing-based function-hiding inner product encryption \cite{kim_ipe_2018} allows computing the inner product of two encrypted vectors. This can be used for simple privacy-preserving data classification tasks, thus enabling a new paradigm in the field of secure computation.

Only special pairing-friendly elliptic curves can be used for pairing-based cryptographic protocols. The security of commonly used BN-254 and BN-256 pairing-friendly curves (based on 254b and 256b prime fields respectively) has been compromised by recent advances in cryptanalysis \cite{ietf_pairingcurves_2020}.
The BLS12-381 pairing-friendly elliptic curve, based on a 381b prime field, has been recently proposed and is part of ongoing standardization led by the Internet Engineering Task Force (IETF) \cite{ietf_pairingcurves_2020}. Along with strong security, the new curve also has higher computational complexity, thus making it challenging to implement on low-power embedded devices.
While previous work \cite{unterluggauer_pairinghw_2014, han_pairinghw_2015, ikeda_pairinghw_2019} implement hardware for 254b and 256b BN curves, efficient hardware accelerators for BLS12-381 are largely unexplored.
In this work, we present a low-power BLS12-381 elliptic curve pairing crypto-processor, and this is the first ASIC implementation supporting the BLS12-381 curve (extended version of \cite{banerjee_cicc_2021}). Our design enables two orders of magnitude energy savings through efficient hardware acceleration, implements countermeasures against timing and power side-channel attacks, and provides flexibility to implement
various ECC and PBC protocols for IoT applications.

\section{BLS12-381 Hardware Implementation}

The pairing computation is a bilinear map $e: \mathbb{G}_1 \times \mathbb{G}_2 \rightarrow \mathbb{G}_T$, where $\mathbb{G}_1$, $\mathbb{G}_2$ are elliptic curve groups and $\mathbb{G}_T$ is a finite field group. This map satisfies the bilinear property $e(aP, bQ) = e(P, Q)^{ab}$, 
where $P \in \mathbb{G}_1$, $Q \in \mathbb{G}_2$, $a, b \in \mathbb{Z}_q^{*}$, $q$ being the group order (a prime). This is the bilinear property 
which enables novel pairing-based cryptographic protocols.
For BLS12-381, the groups $\Ga$ and $\Gb$ are based on elliptic curves $E(\Fp) : y^2 = x^3 + 4$ and $E'(\Fpa) : y^2 = x^3 + 4 \, (1 + \alpha)$ respectively and $\Gt$ is based on the extension field $\Fpd^{*}$, where $p$ is a 381-bit prime and the order $q$ of each group is a 255-bit prime \cite{ietf_pairingcurves_2020}.

%\begin{figure}[!t]
%\centering
%\includegraphics[width=2.5in]{Figures/Towered_Arithmetic.png}
%\caption{Computation stack for towered extension field and elliptic curve arithmetic operations required in BLS12-381 pairings.}
%\label{pairing_compute_stack}
%\end{figure}

\subsection{Finite Field Arithmetic}

Software profiling indicates that big integer prime field arithmetic, especially modular multiplication, accounts for more than 90\% of PBC computation cost. For BLS12-381, we need arithmetic over the 381-bit prime field $\Fp$ (base field) and the 255-bit prime field $\Fq$ (scalar field).

\begin{figure}[!t]
\centering
\includegraphics[width=2.8in]{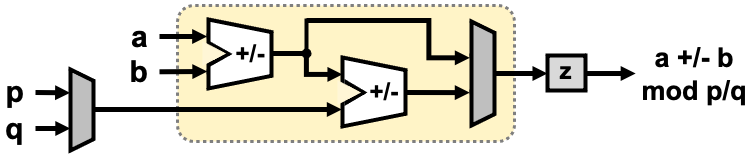}
\caption{Design of modular adder-subtractor for $\Fp$ and $\Fq$.}
\label{pairing_mod_addsub}
\end{figure}

Our modular adder-subtractor design, shown in Fig. \ref{pairing_mod_addsub}, consists of a pair of cascaded 381b adder-subtractors. The modulus can be selected between $p$ and $q$ using a multiplexer. The most significant 126 bits of the data-path are gated when operating over $\Fq$ instead of $\Fp$. Modular reduction is performed using conditional subtraction / addition in the same cycle to avoid timing side-channel leakage.

\begin{figure}[!t]
\centering
\includegraphics[width=3.4in]{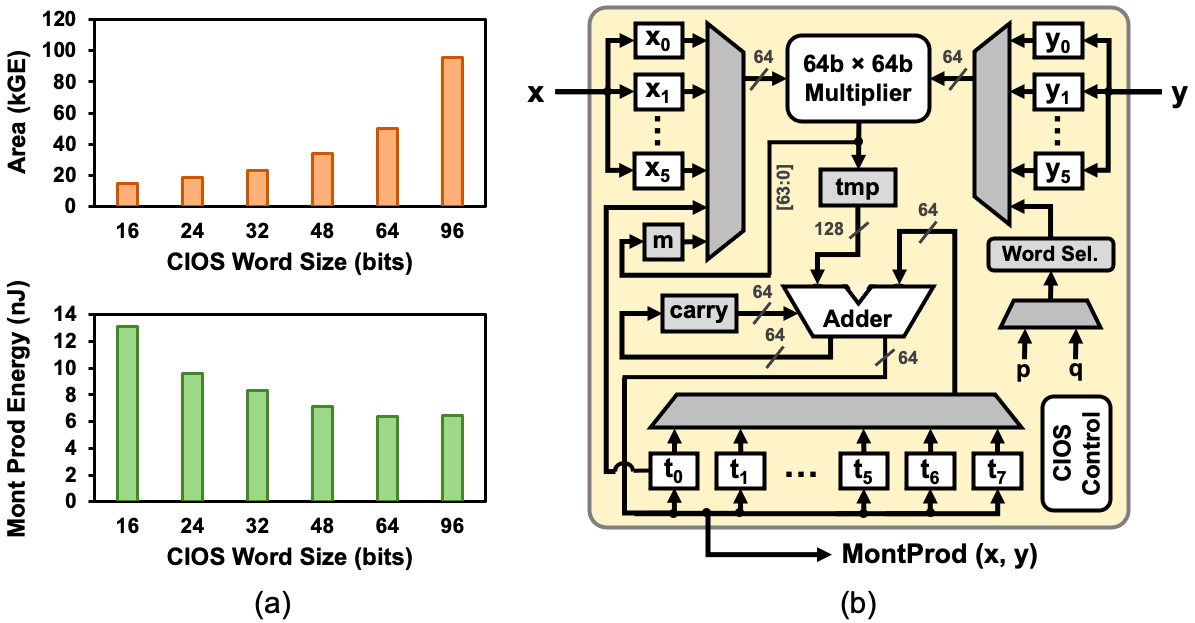}
\caption{(a) Area and energy consumption (measured at 0.66~V) of CIOS Montgomery product in hardware with different word sizes, (b) Architecture of our CIOS Montgomery multiplier with 64-bit words.}
\label{pairing_mod_montmul}
\end{figure}

Montgomery modular multiplication is standard for such large prime fields. Previous work on pairing accelerators use either high-performance parallel pipelined multipliers with large area overhead \cite{han_pairinghw_2015, ikeda_pairinghw_2019} or compact serial multipliers with low energy-efficiency \cite{unterluggauer_pairinghw_2014}. To balance area and energy-efficiency, we implement Montgomery modular multiplication using the coarsely integrated operand scanning (CIOS) approach \cite{koc_montmul_1996}.
Instead of computing multiplication and reduction separately, CIOS performs both operations together in an interleaved manner. Each input is split into $s$ words of size $w$ bits. The core CIOS loop requires $s(2s+1)$ and $2(2s^2 + 2s + 1)$ multiplications and additions respectively, all with $w$-bit word size. The final output needs to be adjusted from modulo $2p$ to modulo $p$ using a conditional subtraction, which is performed in constant time using our single-cycle modular adder.

In order to identify the ideal word size for our application, we have profiled CIOS hardware architectures with word size $w \in \{16, 24, 32, 48, 64, 96\}$ (with $s = \lceil \, 384 / w \, \rceil$). Their area and energy consumption are compared in Fig. \ref{pairing_mod_montmul}a. Clearly, the energy consumption saturates at 64b word size, with 50\% and 25\% lower energy than conventional 16b and 32b architectures respectively. Therefore, we implement CIOS Montgomery multiplication in hardware with $w = 64$ ($\Rightarrow s = 6$), as shown in Fig. \ref{pairing_mod_montmul}b. We split zero-padded inputs into six 64b words and operate on them iteratively using a 64b $\times$ 64b multiplier and a 128b $+$ 64b $+$ 64b adder, both utilizing carry-save structures for shorter critical path delay.

We implement modular inversion using exponentiation following Fermat’s theorem \cite{hankerson_ecc_2006}, and inversion in $\Fp$ and $\Fq$ require 608 and 417 modular multiplications 
(including squarings) 
respectively.

Unlike traditional ECC, pairing computation also requires arithmetic over extensions of the field $\Fp$. For BLS12-381, these are $\Fpa = \Fp[\alpha]/(\alpha^2 + 1)$, $\Fpc = \Fpa[\beta]/(\beta^3 - 1 - \alpha)$ and $\Fpd = \Fpc[\gamma]/(\gamma^2 - \beta)$ \cite{ietf_pairingcurves_2020}.
This construction of the form $\Fp \rightarrow \Fpa \rightarrow \Fpc \rightarrow \Fpd$ is known as towered arithmetic.
Extension field arithmetic over $\Fpa/\Fp$, $\Fpc/\Fpa$ and $\Fpd/\Fpc$ involves manipulation of polynomials with coefficients in $\Fp$. We speed up multiplications, squarings and inversions in these extension fields by extensively using Karatsuba-style divide-and-conquer techniques \cite{mrabet_pairing_2017}, which provides 35\% reduction in pairing energy consumption \cite{banerjee_phdthesis_2021}.

\subsection{Elliptic Curve and Pairing Computations}

We use homogeneous projective coordinates for all elliptic curve point operations. To prevent side-channel vulnerabilities, we employ optimized exception-free point doubling and complete point addition formulas based on \cite{renes_complete_2016}. This ensures that the implementation is constant-time and avoids any data-dependent conditional executions.

Elliptic curve scalar multiplication (ECSM) is implemented using the double-and-add-always algorithm to prevent timing side-channel leakage \cite{fan_ecc_2010}. Our implementation of constant-time ECSM with a 255b scalar requires $4,847 M_1 + 14,025 A_1 + I_1$ (where $I_1 \equiv 608 M_1$) for $\Ga$ and $4,337 M_2 + 510 S_2 + 10,200 A_2 + I_2$ (where $I_2 \equiv 4 M_1 + 2 A_1 + I_1$) for $\Gb$, where $A_1$ (resp. $A_2$), $M_1$ (resp. $M_2$) and $I_1$ (resp. $I_2$) denote additions / subtractions, multiplications / squarings and inversions respectively in $\Fp$ (resp. $\Fpa$). ECSM performance can be improved by using standard pre-computation techniques (memory-time trade-offs) such as windowing, comb, etc \cite{hankerson_ecc_2006}.

The two main components of pairing computation are Miller Loop (ML) and Final Exponentiation (FE) \cite{mrabet_pairing_2017}.
The Miller Loop consists of a series of line computations based on binary representation of the curve parameter $u$.
For our implementation of BLS12-381, their computation costs, in terms of equivalent number of $\Fp$ multiplications, are $7,050 M_1$ and $8,339 M_1$ respectively.
The overall pairing involves $15,389$ $\Fp$ multiplications and requires $\approx$3.4M cycles.

Many practical pairing-based protocols require evaluating the product of several pairings, also known as multi-pairing:
%$\prod_{j=1}^{n} \, e(P_j, Q_j) \, = \, e(P_1, Q_1) \times e(P_2, Q_2) \times \, \cdots \, \times e(P_n, Q_n)$.
\[
\prod_{j=1}^{n} \, e(P_j, Q_j) \, = \, e(P_1, Q_1) \times e(P_2, Q_2) \times \, \cdots \, \times e(P_n, Q_n)
\]
We share ML and FE computations across multiple pairing instances to speed up multi-pairing, which is especially useful in aggregate signature verification \cite{boneh_blockchain_2018}. Sharing only the FE operation provides $2 \times$ improvement, while sharing both ML and FE provides another $30\%$ energy savings \cite{banerjee_phdthesis_2021}.

Function-hiding inner product encryption \cite{kim_ipe_2018} requires scalar multiplications on $\Gb$. The corresponding twist curve $E'$ supports the skew Frobenius map \cite{iijima_frobeniusmap_2002}, which allows efficient computation of $\hat{\phi}(P) = pP$ $\forall \, P \in E'(\Fpa)$. To speed up the encryption step, we split the 255-bit scalar $k < q$ into two 128-bit parts as $k = k_1 + k_2 u^2$ where $p \equiv u$ (mod $q$) and compute $kP = k_1 P + k_2 \, (u^2 P) = k_1 P + k_2 \, \hat{\phi}( \hat{\phi}(P) )$ using multi-exponentiation \cite{hankerson_ecc_2006}. This leads to $1.8 \times$ energy savings in inner product encryption per vector element \cite{banerjee_phdthesis_2021}.

Pairing-based signatures also require mapping random $\Fp$ elements to points on elliptic curves, known as the hash-to-curve operation \cite{boneh_blockchain_2018}. Our implementation of hash-to-$\Ga$ requires 1,897 $\Fp$ multiplications. Since our modular arithmetic unit can be configured for the 255-bit prime $q$, we also support ECC using Jubjub, a twisted Edwards curve over $\Fq$ \cite{ietf_pairingcurves_2020}. Our constant-time Jubjub ECSM requires 4,755 $\Fq$ multiplications.

\section{Pairing Crypto-Processor Architecture}

\begin{figure}[!t]
\centering
\includegraphics[width=3.4in]{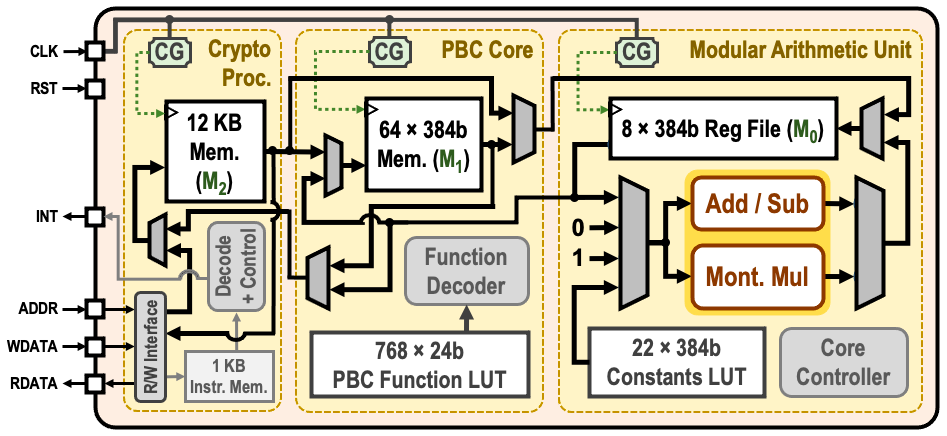}
\caption{Top-level architecture of our BLS12-381 pairing crypto-processor.}
\label{pairing_core_arch}
\end{figure}

The top-level architecture of our pairing crypto-processor is shown in Fig. \ref{pairing_core_arch}, where the arithmetic unit is integrated with a 15.375~KB data memory, a 1~KB instruction memory and an instruction decoder. It can be programmed using 32b custom instructions to perform different modular arithmetic, elliptic curve and pairing computations.

%\clearpage

\begin{figure}[!t]
\centering
\includegraphics[width=3.2in]{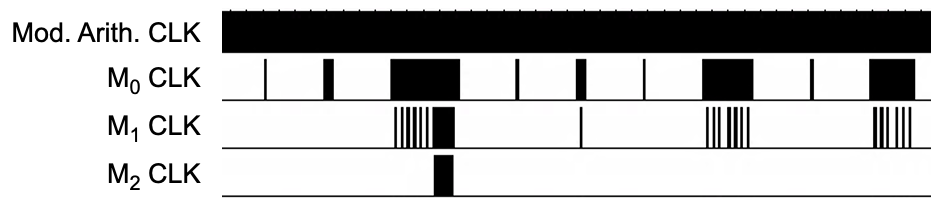}
\caption{Simulated waveforms showing memory clock gating in pairing crypto-processor during a snapshot of the final exponentiation computation.}
\label{pairing_mem_clk_gate}
\end{figure}

The crypto-processor data memory is hierarchical with three levels. First, the modular arithmetic unit is coupled with a small 8 $\times$ 384-bit register file M\textsubscript{0} implemented using flip-flops for efficiency. M\textsubscript{0} is used for all $\Fp$ and $\Fpa$ arithmetic computations. At the next level, a 64 $\times$ 384-bit SRAM M\textsubscript{1} is used to store all temporary variables required for $\Fpc$, $\Fpd$, $\Ga$, $\Gb$ and $\Gt$ arithmetic computations. Finally, a 256 $\times$ 384-bit SRAM M\textsubscript{2} is used to store protocol-level inputs and outputs.
While simple ECSM and pairing computations require only few of these 256 memory locations in M\textsubscript{2}, having a large top-level memory is useful to support efficient multi-pairings and hash-to-curve maps.
Each memory module is dynamically clock gated based on the function under execution, providing up to 20\% power savings. Fig. \ref{pairing_mem_clk_gate} shows simulated waveforms for memory clock gating during final exponentiation computation.
The modular arithmetic unit is operational continuously, while the memories are accessed only when data movement is required.

The pairing computation requires several constants in Montgomery domain, which are stored in a 22 $\times$ 384-bit lookup table (LUT).
The $\Fp$ and $\Fpa$ functions are handled by the modular arithmetic unit.
Optimized micro-code for $\Fpc$, $\Fpd$, $\Ga$, $\Gb$ and $\Gt$ arithmetic functions
(which require $\Fp$ and $\Fpa$ arithmetic)
are stored in another 768 $\times$ 24-bit LUT. To save area, these LUTs are realized entirely using digital logic. Combined area of these two LUTs is only 6k-gate, which is 53k-gate and 34k-gate smaller than SRAM-based and ROM-based implementations respectively.

\section{System Design and Measurement Results}

\subsection{Chip Architecture and Test Setup}

\begin{figure}[!t]
\centering
\includegraphics[width=3.4in]{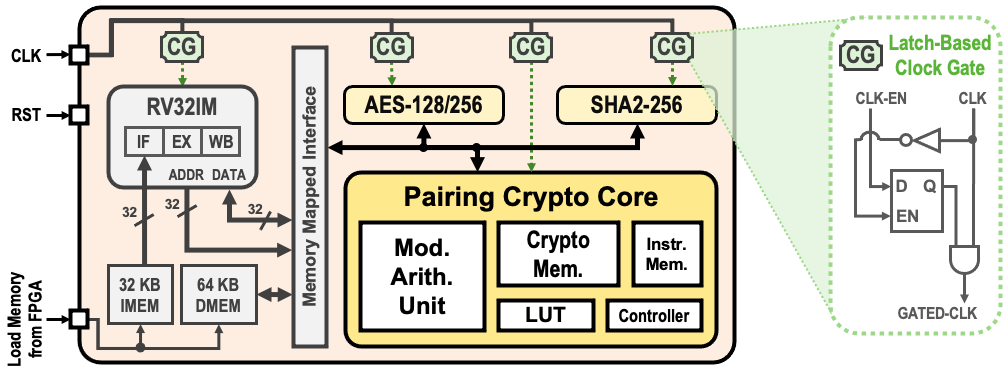}
\caption{Chip architecture with pairing core and RISC-V micro-processor.}
\label{pairing_chip_arch}
\end{figure}

As shown in Fig. \ref{pairing_chip_arch}, the pairing crypto-processor is integrated (through a memory-mapped interface) with a low-power RISC-V micro-processor supporting the RV32IM instruction set, with 1-cycle multiplier, 32-cycle divider, 32~KB instruction memory and 64~KB data memory \cite{banerjee_tches_2019}.
The RISC-V is also coupled with accelerators for AES-128/256 and SHA2-256 \cite{banerjee_jssc_2019}.
Reading from and writing to accelerators through the memory-mapped interface require 3 cycles and 2 cycles respectively.
The RISC-V, AES, SHA and pairing cores all have dedicated clock gates independently configurable for power savings. The RISC-V core can be clock-gated using wait-for-interrupt instruction, and it is woken up by dedicated interrupts from the cryptographic accelerators.
The accelerators can be accessed through software using simple load and store instructions, without any changes to the compilation toolchain.
The RISC-V processor is used for scheduling cryptographic workloads as well as for processing their inputs and outputs.

\begin{figure}[!t]
\centering
\includegraphics[width=3.4in]{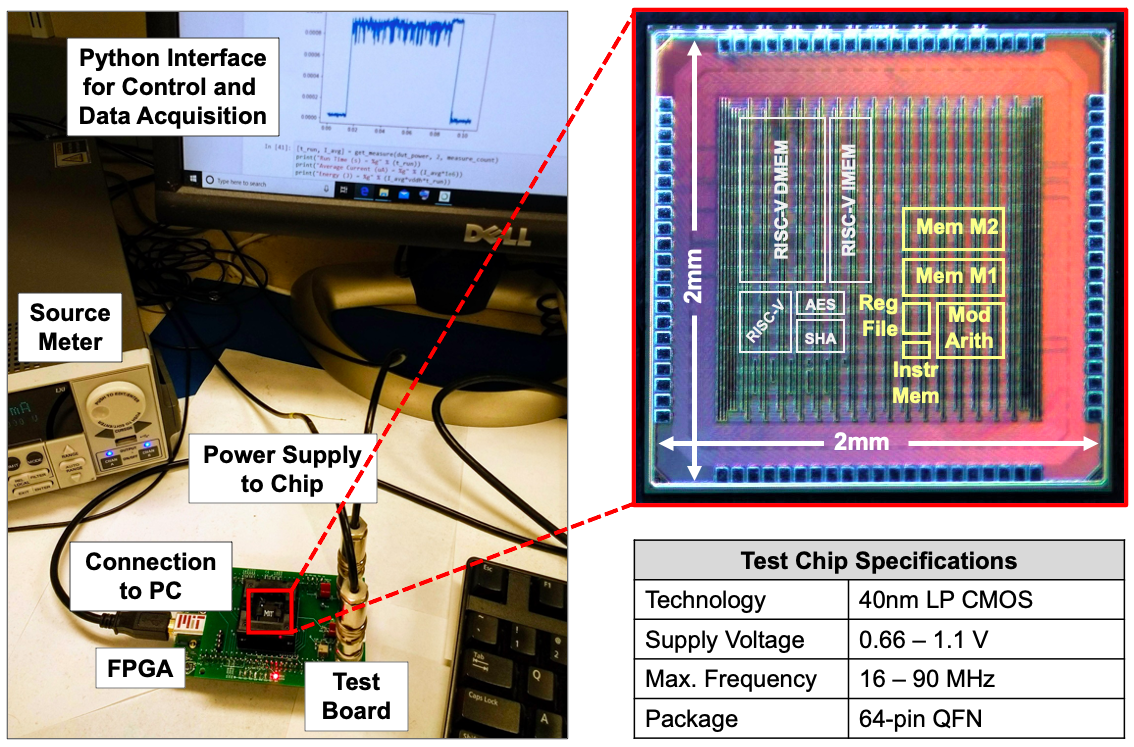}
\caption{Measurement setup and chip micrograph.}
\label{pairing_test_setup}
\end{figure}

\begin{figure}[!t]
\centering
\includegraphics[width=3.4in]{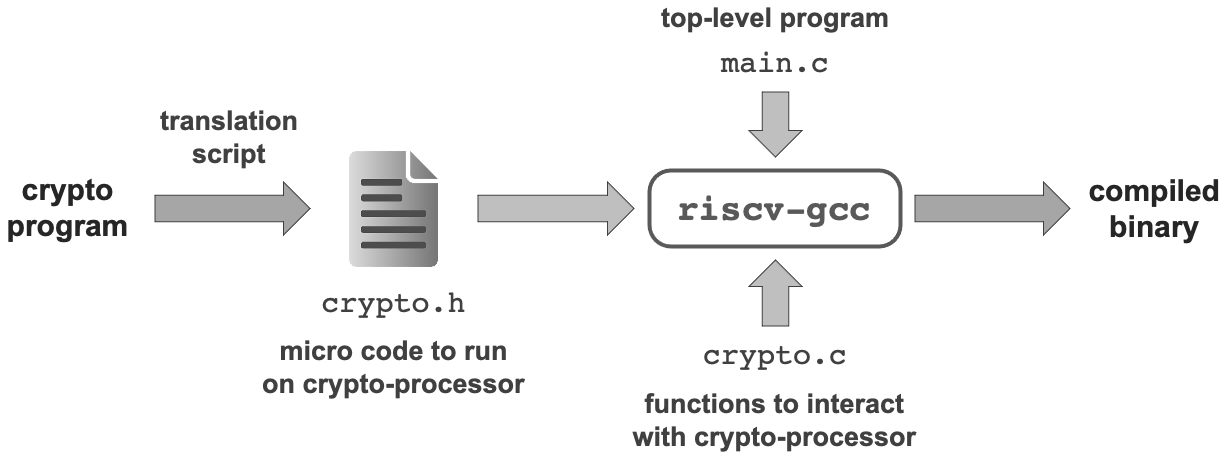}
\caption{Program compilation and integration with RISC-V software flow.}
\label{program_compilation}
\end{figure}

\begin{figure}[!t]
\centering
\includegraphics[width=3.4in]{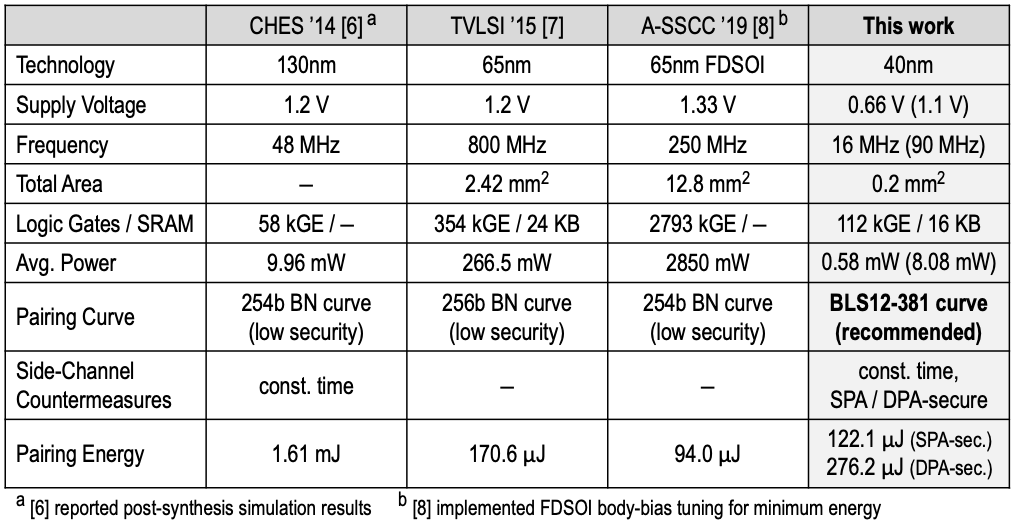}
\caption{Comparison with previous work on pairing hardware accelerators.}
\label{comparison_table}
\end{figure}

Our test chip (Fig. \ref{pairing_test_setup}) was fabricated in the TSMC 40nm low-power process.
Our pairing crypto-processor consists of 112k logic gates and 16 KB SRAM, with a total area of 0.2 mm$^2$ (logic and memory combined). The RISC-V (including interrupt controller and peripherals), AES and SHA cores occupy 52k, 12k and 23k logic gates respectively.
Our test chip supports supply voltage scaling from 0.66~V to 1.1~V, and its maximum operating frequency (for both RISC-V and accelerators) at 0.66~V and 1.1~V are 16~MHz and 90~MHz respectively.
Fig. \ref{pairing_test_setup} shows our measurement setup. An FPGA is used to transfer instructions / programs from a host computer to the instruction memory of our test chip. The crypto-processor programs are translated into appropriate format using a Python script, which is then integrated together with RISC-V software.

Fig. \ref{pairing_test_setup} shows our measurement setup with the test chip housed in a QFN64 socket on a custom board, an Opal Kelly XEM7001 FPGA interfaces with the chip, and a Keithley 2602A source meter is used to supply power. The FPGA is used to transfer instructions / programs from a host computer to the instruction memory of our test chip. All elliptic curve and pairing-based cryptography programs are written using custom instructions and compiled with a Python script, which is then integrated together with the RISC-V software (Fig. \ref{program_compilation}).

Fig. \ref{comparison_table} compares our design with previous work on pairing hardware accelerators. We are the first to demonstrate the higher security BLS12-381 curve in hardware. Our design is an order of magnitude more energy-efficient than the embedded-scale accelerator in \cite{unterluggauer_pairinghw_2014}. Compared to the high-performance accelerators in \cite{han_pairinghw_2015, ikeda_pairinghw_2019}, our design is an order of magnitude smaller with significantly lower power consumption. We also 
achieve the lowest area-energy product compared to previous designs, 
implement side-channel countermeasures for stronger security, and have the flexibility to accelerate signature aggregation, functional encryption and other PBC protocols in hardware.

\subsection{Protocol Implementation Results}

\begin{figure}[!t]
\centering
\includegraphics[width=3.4in]{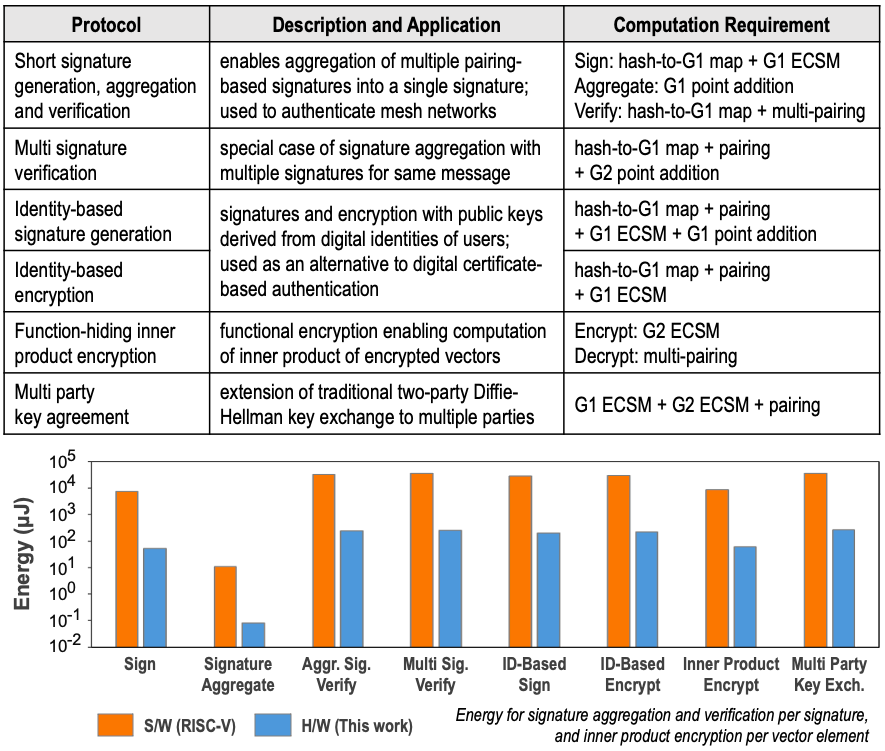}
\caption{Pairing-based cryptography protocol implementation benchmarks.}
\label{pairing_protocol_benchmarks}
\end{figure}

\begin{figure}[!t]
\centering
\includegraphics[width=2.7in]{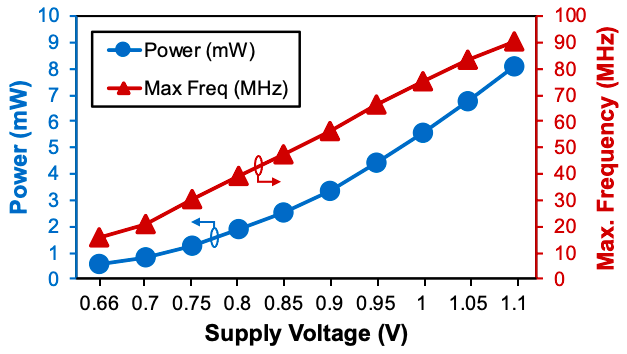}
\caption{Measurement of voltage-frequency scaling of our test chip.}
\label{voltage_scaling}
\end{figure}

To measure the efficiency of our design as well as to demonstrate its flexibility in supporting various security applications, we have implemented and profiled several BLS12-381 pairing-based cryptography protocols on our test chip, as detailed in Fig. \ref{pairing_protocol_benchmarks}. Our hardware-accelerated implementations are 130-140$\times$ more energy-efficient compared to software. All measurement results are reported at 16~MHz and 0.66~V, the operating condition providing lowest energy consumption. Effect of voltage-frequency scaling is shown in Fig. \ref{voltage_scaling}.

\subsection{Side-Channel Countermeasures}

\begin{figure}[!t]
\centering
\includegraphics[width=2.7in]{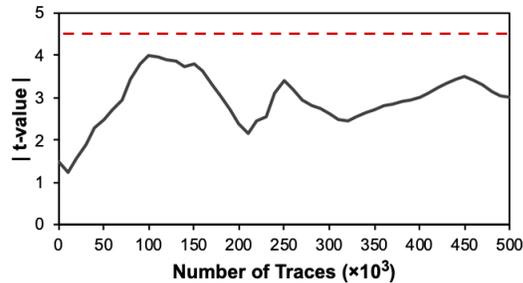}
\caption{Side-channel leakage test with SPA and DPA countermeasures.}
\label{leakage_test}
\end{figure}

As countermeasures against timing and simple power analysis (SPA) side-channel attacks \cite{fan_ecc_2010}, we use complete point addition formulas \cite{renes_complete_2016} and double-and-add-always technique \cite{coron_eccdpa_1999} in our ECSM and pairing implementations.
All results discussed earlier are from constant-time implementations with these countermeasures. To prevent stronger differential power analysis (DPA) attacks, we employ following countermeasures:
\begin{itemize}
\itemsep0em
\item randomized projective coordinates \cite{cietjoye_eccdpa_2003, vercauteren_pairingsca_2004}, where elliptic curve points $(X : Y : Z)$ are transformed into the form $(\lambda X : \lambda Y : \lambda Z)$ with non-zero random $\lambda \in \Fp$
\item ECSM with random scalar splitting \cite{coron_eccdpa_1999}, where secret scalar $k \in \Fq$ is split into two parts $r$ and $k-r$ with random $r \in \Fq$, computed as $kP = rP + (k-r)P$
\item pairing with random exponents and bilinear property \cite{vercauteren_pairingsca_2004}, computed as $e(aP, bQ)$ = $e(P, Q)^{ab}$ = $e(P,Q)$ with random $a\in \Fq$ and $b = a^{-1}$ mod $q$
\end{itemize}
The first technique is practically free, requiring only a few $\Fp$ multiplications. The second technique can be significantly simplified by using multi-exponentiation \cite{hankerson_ecc_2006}, where $2P$ is pre-computed and both scalars $r$ and $k-r$ are processed simultaneously to share point doublings and merge point additions. The third technique is quite expensive, requiring one $\Ga$ ECSM, one $\Gb$ ECSM and one $\Fq$ inversion. We also require generation of random elements in $\Fp$ or $\Fq$, performed by using our SHA2-256 accelerator as a cryptographically secure pseudo-random number generator.
Fig. \ref{leakage_test} shows measured power side-channel leakage assessment results (based on standard non-specific fixed vs. random $t$-test \cite{schneider_leakage_2015}) over 0.5M traces for our hardware implementation with all countermeasures. Here, $|t\text{-value}| < 4.5$ indicates, with very high confidence, that there is no evidence of first-order side-channel leakage.
We have also performed difference-of-means test to verify that there is no significant observable information leakage through power consumption for any bit in ECSM secret scalar being 0 versus 1.
For ECSM, the DPA countermeasures have only 10\% performance and energy overheads. However, DPA countermeasures lead to $2.3 \times$ additional overhead for pairing.
We note that all these countermeasures can be implemented without any changes to our hardware, by utilizing the programmability of our crypto-processor.

\section{Conclusion}
In this work, we have presented a low-power programmable crypto-processor to accelerate ECC and PBC using the recently proposed BLS12-381 pairing-friendly elliptic curve. Using optimized algorithms and low-power area-efficient architectures, we demonstrate practical hardware-accelerated pairings which enable novel cryptographic protocols to secure resource-constrained IoT devices.

\section*{Acknowledgment}
%The authors thank Texas Instruments for funding and TSMC University Shuttle Program for chip fabrication support.
The authors would like to thank Texas Instruments for funding this work, the TSMC University Shuttle Program for chip fabrication support, and Bluespec, Xilinx, Cadence, Synopsys and Mentor Graphics for providing CAD tool support.

%\balance
\bibliographystyle{IEEEtran}
\bibliography{references_preprint}

%\clearpage

\end{document}